\documentclass[10pt]{iopart}

\usepackage{iopams}  

\usepackage{graphicx}
\usepackage{siunitx}
\usepackage{floatrow}
\usepackage{relsize}
\usepackage{gensymb}
\usepackage{xcolor}
\usepackage[square,numbers,sort&compress]{natbib}

\begin{document}

\title{Gate control of 2D magnetism in tri- and four-layers $\rm CrI_3$/graphene heterostructures}

\author{Ping Wang$^{1,3}$, Fuzhuo Lian$^{1,3}$, Renjun Du$^{1,3}$, Xiaofan Cai$^1$, Song Bao$^1$, Yaqing Han$^1$, Jingkuan Xiao$^1$, Kenji Watanabe$^2$, Takashi Taniguchi$^2$, Jinsheng Wen$^1$, Hongxin Yang$^1$, Alexander S. Mayorov$^{1,*}$, Lei Wang$^{1,*}$ and Geliang Yu$^{1,*}$ }

\address{$^1$ National Laboratory of Solid State Microstructures, School of Physics, Nanjing University, Nanjing 210093, China}
\address{$^2$ National Institute for Material Science, 1-1 Namiki, Tsukuba 305-0044, Japan}
\address{$^3$ These authors contributed equally to this work.}
\address{$^*$ Authors to whom any correspondence should be addressed.}

\ead{leiwang@nju.edu.cn and yugeliang@nju.edu.cn}

\vspace{10pt}
\begin{indented}
	\item[]\today
\end{indented}
\vspace{10pt}

\begin{abstract}

\section*{}
\label{abstract}
We conduct experimental studies on the electrical transport properties of monolayer graphene directly covered by a few layers of $\rm CrI_3$.  We do not observe the expected magnetic exchange coupling in the graphene but instead discover proximity effects featuring gate and magnetic field tunability. The tunability of gate voltage is manifested in the alignment of the lowest conduction band of $\rm CrI_3$ and the Fermi level of graphene, which can be controlled by the gate voltage. The coexistence of the normal and atypical quantum Hall effects in our device also corresponds to gate-control modulation doping. The lowest conduction band depends on the magnetic states of the $\rm CrI_3$ and can be altered by the magnetic field, which corresponds to the resistance loops during back-and-forth sweeps of the magnetic field. Our results serve as a reference for exploiting the magnetic proximity effects in graphene.
\end{abstract}

\ioptwocol


\section*{}
\label{introduction}
Van der Waals heterostructures provide a method for constructing an artificial interface system, allowing different materials to be efficiently stacked together in a predetermined order \cite{Liu2016NatRevMaterVanWaalsHeterostructures,Hellman2017Rev.Mod.Phys.InterfaceinducedPhenomenaMagnetism}. The atomic-level flatness of the hexagonal boron nitride (hBN) surface and minimal interface Coulomb coupling allows graphene encapsulated in hBN to closely approach theoretical expectations\cite{Dean2010NatureNanotechBoronNitrideSubstrates,Mayorov2011NanoLett.MicrometerScaleBallisticTransport}. By replacing hBN with other two-dimensional materials, a diverse range of proximity systems can be obtained\cite{Zutic2019MaterialsTodayProximitizedMaterials,Burch2018NatureMagnetismTwodimensionalVan}. Chromium iodide ($\rm CrI_3$)\cite{Huang2017NatureLayerdependentFerromagnetismVan}, as a two-dimensional magnetic material, has been extensively studied for its unique magnetic properties, including stacking dependence\cite{Sivadas2018NanoLett.StackingDependentMagnetismBilayer,Xu2022Nat.Nanotechnol.CoexistingFerromagneticAntiferromagnetic}, electrical/doping control\cite{Jiang2018NatureNanotechControllingMagnetism2D,Huang2018NatureNanotechElectricalControl2Da,Marian2023npj2DMaterApplElectricallyTunableLateral}, optical control\cite{Dabrowski2022NatCommunAllopticalControlSpin,Cheng2021NatCommunLightHelicityDetector}, and strain/pressure control\cite{Wu2019Phys.Chem.Chem.Phys.StraintunableMagneticElectronic,Huang2023NatCommunPressurecontrolledMagnetism2D}. Theoretical calculations suggest that the magnetic proximity system of $\rm CrI_3$ and monolayer graphene (MLG) is expected to achieve the anomalous Hall effect\cite{Zhang2018Phys.Rev.BStrongMagnetizationChern,Ren2023AdvancedPhysicsResearchTunableQuantumAnomalous}.

The biggest challenge in the experimental study of $\rm CrI_3$/MLG is the extreme air sensitivity of $\rm CrI_3$\cite{Zhang2022J.Am.Chem.Soc.DegradationChemistryKinetic,Shcherbakov2018NanoLett.RamanSpectroscopyPhotocatalytic}, while electric transport devices inevitably require the preparation of ohmic contact electrodes. To protect $\rm CrI_3$ from damage as much as possible, we used pre-fabricated Pt electrodes on a hBN substrate and a dry-transfer method\cite{wangOneDimensionalElectricalContact2013} to stack the flakes in a glovebox (filled with argon (Ar), and $\rm H_2O, O_2<$ 0.1 ppm). After the transfer process, the Polycarbonate (PC) film used in the transfer process can take advantage of the excellent low-temperature characteristics of PMMA (950K A2), avoiding the use of chloroform to remove the PC which is not safe for $\rm CrI_3$\cite{Zhao2020Nat.Mater.MagneticProximityNonreciprocal}. Please refer to the Supporting Information (S.I. Figs S1 and S2) for more details. Therefore, throughout the entire process, $\rm CrI_3$ is not exposed to air. In addition to the protection provided by hBN, the PC+PMMA film also isolates the sample from the air, providing an additional layer of protection\cite{Tang2020AdvancedMaterialsMagneticProximityEffect}.

Fig 1a shows the schematic of the general device structure we fabricate (for information on the other three devices, please refer to S.I. Fig S3). Monolayer graphene, which has an elongated rectangular shape is fully covered by $\rm CrI_3$ and encapsulated by hBN. It comes into contact with the pre-patterned Pt electrodes on the bottom layer of hBN. The orientation of the graphene stripe is arranged to be as parallel as possible to the pre-patterned Hall channel. This minimizes the mixing of longitudinal resistance ($R_{\rm xx}$) and transverse resistance ($R_{\rm xy}$). The top gate is graphite, and the bottom gate is made of p-doped Si,  serving as the top gate voltage ($V_{\rm tg}$) and the bottom gate voltage ($V_{\rm bg}$), respectively. Figure 1b displays the optical image of  D1 featuring a $\rm CrI_3$ flake with three and four layers in thickness labeled as 3L and 4L (inset: an optical image of $\rm CrI_3$  before stacking). The boundary of this 1L thickness step is positioned between two pairs of Pt electrodes. However, we did not find that the thickness of $\rm CrI_3$ significantly impacts the main results. The different thicknesses only result in variations in the critical magnetic field\cite{Huang2017NatureLayerdependentFerromagnetismVan} $ B_{\rm c}$,  where 2L exhibits $|B_{\rm c}|\approx$0.8 T, while thickness greater or equal to 3L displays $|B_{\rm c}|\approx$1.8 T.

\begin{figure*}[htb]

\includegraphics[scale=0.5]{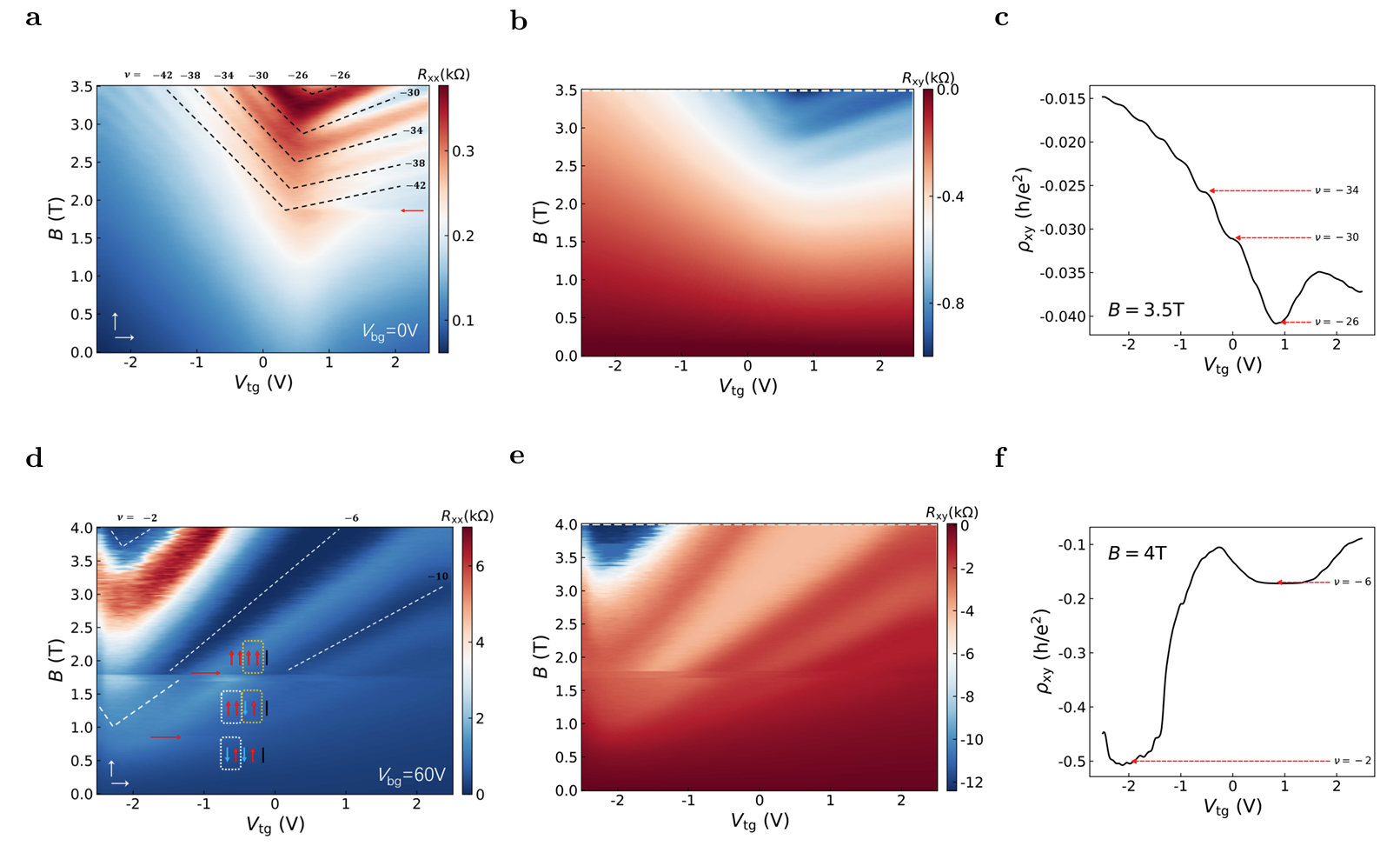}
\caption{\textbf{Landau fan diagrams of \textbf{$\rm CrI_3$}/MLG.} \textbf{a-b.} Longitudinal $R_{\rm xx}$ and Hall $R_{\rm xy}$ resistance of D1 4L region acquired by sweeping $V_{\rm tg}$ and $B$ according to the direction indicated by the white arrows with $V_{\rm bg}$=0 V, respectively. \textbf{c.} $\rho_{\rm xx}$ cuts from panel b with $B=$3.5 T; \textbf{d-f.} Identical to \textbf{a-c} but with $V_{\rm bg}$=60 V.}
\end{figure*}

\label{Fig 1cd }
 
Fig 1c (d) shows the four-terminal resistance at the 4L region obtained by sweeping  $V_{\rm tg}$ ($V_{\rm bg}$) back and forth while grounding the other gate. This was conducted with $T=1.6$ K and $B=0$ T. The observed hysteresis in resistance during the back-and-forth sweep indicates a substantial charge transfer\cite{Tseng2022NanoLett.GateTunableProximityEffects} ($\approx 4.5\times 10^{12}\rm cm^{-2}$, as shown in Fig S4) from graphene to $\rm CrI_3$. Inserting a few hBN layers between the $\rm CrI_3$ and the MLG as a spacer will suppress the charge transfer as shown in Fig S5 in D3\cite{Britnell2012NanoLett.ElectronTunnelingUltrathin,Lee2011Appl.Phys.Lett.ElectronTunnelingAtomically}. The resistance peak around 0 V is not the charge neutrality or the Dirac point (DP). Based on the measurement of Hall resistance (Fig 2b and 2e), graphene is consistently hole-doped throughout the gate range. Here, we label this anomalous peak as $R_{\rm AP}$ ($V_{\rm AP}$ is the corresponding gate voltage) and will explain its origin later.

Fig 1e shows the resistance map obtained by sweeping both gates in the direction indicated by the white arrows at $B=0$ T. The movement of $ R_{\rm AP}$ is similar to the Dirac points moving along the displacement field in the dual-gate structures. However, it does not move in a straight line as Dirac points, as can be seen more clearly in Fig 1f, obtained from the same parameter space as Fig 1e at $B=4$ T. The white dashed line represents the trajectory of $R_{\rm AP}$. The map can be divided into two clearly distinct regions by $R_{\rm AP}$, labeled as I and II. Region I presents classical graphene magneto-transport, characterized by a series of parallel trajectories (yellow dashed lines) with a negative slope and the spacing denoted as $\Delta_{\rm I}$. The negative slope is consistent with the ratio of the geometrical capacitances of the top and bottom gates. The straight line indicates that the carrier concentration ($n_{\rm g}$) can be determined using the standard expression $n_{\rm g}=(C_{\rm tg}V_{\rm tg}+C_{\rm bg}V_{\rm bg})/e-n_0$. Here, $C_{\rm tg}$ and $C_{\rm bg}$ are the top and bottom gate capacitances per area, respectively, $e$ is the elementary charge, $n_0$ is the residual doping\cite{Maher2014ScienceTunableFractionalQuantum}. In contrast, after $V_{\rm AP}$, these trajectories suddenly reverse from a negative slope to a positive slope, deviating from the straight line, and the spacing  $\Delta_{\rm II}$ (red dashed line) also increases.

Region II corresponds to the hysteresis region observed in the back-and-forth gate sweeping shown in Fig 1c and 1d. Therefore, the difference between regions I and II shares the same source of the gate hysteresis. At the interface of $\rm{CrI_3}$/MLG, due to the electron affinity of $\rm{CrI_3}$ and the work function of graphene, electrons transfer from graphene to $\rm{CrI_3}$, leaving the $p_{z}$ orbital states of  carbon unoccupied. To fill the vacant $p_{z}$ orbital states, the conduction band of $\rm{CrI_3}$ shifts towards the Fermi level, where the $p_{z}$ orbital states are located\cite{Ren2023AdvancedPhysicsResearchTunableQuantumAnomalous}, as shown in Fig 4a.

The presence of $\rm{CrI_3}$ between MLG and the top gate results in the inequivalent $V_{\rm tg}$ and $V_{\rm bg}$. Only $R_{\rm AP}$ is present in $R(V_{\rm tg})$, and as $V_{\rm tg}$ continues to increase in the positive direction, the resistance dose not increase but continuously decreases. However, in $R(V_{\rm bg})$, after $R_{\rm AP}$, the resistance significantly increases when $V_{\rm bg}$ is increased in the positive direction. This indicates that only $V_{\rm bg}$ can adjust the Fermi level closer to the Dirac point. This is because the electric field generated by $V_{\rm tg}$ must cross the $\rm{CrI_3}$ flake, directly affecting the conduction band's position of $\rm{CrI_3}$\cite{Marian2023npj2DMaterApplElectricallyTunableLateral,Ren2023AdvancedPhysicsResearchTunableQuantumAnomalous}.  Negative $V_{\rm tg}$ causes an upward shift of the conduction band, bringing it above the Dirac point, as shown in Fig 4b. However, positive $V_{\rm tg}$ causes a downward shift, even below the Fermi level of graphene, resulting in more electrons transferring to $\rm{CrI_3}$, as depicted in Fig 4c. Therefore, when $V_{\rm tg}$ is increased from negative to positive, once it reaches $V_{\rm AP}$, charge transfer is enhanced, leading to an increase in MLG hole carriers and a decrease in resistance. Thus, $R_{\rm AP}$ is the product of gate-tunable band alignment (the conduction band of $\rm CrI_3$ and Fermi energy of MLG). Moreover, the screening effect of the electrons transferred into $\rm{CrI_3}$ makes graphene insensitive to $V_{\rm tg}$. Based on the analysis above, to reach the Dirac point, it is necessary to use $V_{\rm tg}$ to raise the conduction band of $\rm{CrI_3}$ above the DP and then use  $V_{\rm bg}$  to adjust the Fermi energy to reach the DP. Therefore, the true DP appears in the bottom right corner of the map. The filling factor $\nu=-2$ in the bottom right corner of Fig 1f indicates that it is very close to the DP.

After $V_{\rm AP}$, the conduction band of $\rm{CrI_3}$ becomes occupied with electrons. However, these electrons do not significantly contribute to electrical conductivity due to their large effective mass, resulting in very low mobility\cite{Tseng2022NanoLett.GateTunableProximityEffects,Tenasini2022NanoLett.BandGapOpening,wangQuantumHallPhase2022b,Yang2023NatCommunUnconventionalCorrelatedInsulator}. The deviation of the trace of $R_{\rm AP}$ and the trajectories in region II from a straight line is attributed to the gate-dependent charge density in $\rm CrI_3$, which does not vary linearly. Shown in Fig 4b, $\Delta_{\rm 1}$ is the energy difference between the graphene Dirac point and the lowest-energy conduction band of the $\rm CrI_3$, which is fixed by the combination of the graphene work function and $\rm CrI_3$ electron affinity. That's to say, $\Delta_{\rm 1}$ also is gate-dependent. Although the trajectory in the rightmost of region II has a positive slope, but not vertical; whereas the trajectories on the left side are almost parallel to $V_{\rm tg}$.  The reason for this difference is the varying density of states in graphene. The closer to the Dirac point, the smaller the density of states in graphene. This means a larger change in chemical potential is required to add electrons in graphene, leading to more electrons tunneling into the $\rm CrI_3$. So, the value  of $\Delta_{\rm II}$ reaches its maximum near the Dirac point as shown in Fig 1f. We think the fact that $\Delta_{\rm II}>\Delta_{\rm I}$ is for the $V_{\rm bg}$ will simultaneously add electrons into both the $\rm CrI_3$ and the graphene. As a result, in region II, the graphene shows smaller $V_{\rm bg}$-induced changes in resistance than in region I.

From the above analysis, it can be observed that the charge transfer in this system can be controlled by gating and, in some cases is  suppressed completely. Next, we will investigate the differences between regions I and II through the Landau fan diagram.

Fig 2a shows the Landau fan diagram acquired by sweeping $V_{\rm tg}$ at $V_{\rm bg}$=0 V. The filling factors are extracted from the corresponding Hall resistivity (Fig 2b).  Fig 2c is a cross-section from Fig 2b at $B=$ 3.5 T. All the filling factors are negative and differ by 4, indicating that MLG is still hole-doped, and the spin- and valley-degeneracy are maintained. $R_{\rm AP}$  appears when $V_{\rm tg}$ is approximately 0.3 V, and to the left of  $R_{\rm AP}$  belongs to region I, while the right belongs to region II. Meanwhile, at $B\approx$1.8 T, there is a significant shift in resistance (indicated by the red arrow), and this shift almost exclusively occurs in region II.

Fig 2d shows the Landau fan diagram obtained by fixing $V_{\rm bg}$=60 V, in which nearly the entire $V_{\rm tg}$ range belongs to region II. Fig 2e shows the corresponding Hall resistance, and  Fig 2f  is a cross-section from Fig 2e  at $B=$ 4 T. $V_{\rm AP}$ is shifted from 0.3 V to -2.3 V, which is consistent with Fig 1e and 1f. In addition to the more pronounced filling factor separation, we observe that the resistance shift at $B\approx$1.8 T is nearly pervasive throughout the entire map (as indicated by the horizontal red arrow in Fig 2d). Furthermore, there is also a weaker shift at $B\approx$0.8 T.

For vertical samples, tunneling between the Landau levels of two graphene layers with a ferromagnetic insulator barrier ($\rm CrBr_3$)  also gives rise to peaks in conductance. These peaks exhibit a distinct shift at the critical temperature ($\rm T_c$), suggesting that the tunneling mechanism between the Landau levels of the two graphene layers is temperature-dependent. Below $\rm T_c$, the dominant mechanism for tunneling is magneto-assisted inelastic tunneling. Above $\rm T_c$, the dominant mechanism switches to phonon-assisted elastic tunneling\cite{Ghazaryan2018NatElectronMagnonassistedTunnellingVan}.  For our case, we can understand the resistance shifts observed in MLG at the critical magnetic field by considering the switching of the interlayer magnetic configuration of $\rm{CrI_3}$ and the corresponding band structure alteration. $\rm{CrI_3}$ exhibits ferromagnetic (FM) coupling for intralayer and antiferromagnetic (AFM) coupling for interlayer. When the interlayer coupling switches from AFM to FM states, the conduction band of $\rm{CrI_3}$ shifts downwards. Fig 4c shows the lowest conduction band difference between AFM and FM states for $\rm{CrI_3}$, denoted as $\Delta_2$. Therefore, when the external magnetic field causes the magnetic configuration of $\rm{CrI_3}$ to switch into the FM state, $\Delta_2$  facilitates the transfer of more electrons to $\rm{CrI_3}$. This will result in a modification of the density of charge carriers in MLG, i.e., the Fermi energy shifts downward. Due to the abrupt change in carrier density in MLG, a resistance shift is measured at the critical magnetic field of $\rm{CrI_3}$. We also notice that the local electronic properties of $\rm{CrI_3}$ have been probed by scanning tunneling microscopy/spectroscopy (STM/STS), revealing that the electronic hybridization between graphene and $\rm{CrI_3}$ dependents on the different interlayer magnetic coupling\cite{Qiu2021NatCommunVisualizingAtomicStructure}.

The shift at $|B|\approx0.8$ T is caused by the interlayer coupling between the topmost layer of 4L $\rm CrI_3$  switching from AFM to FM states. As indicated by the arrows in the vertical direction in Fig 2d, the black line on the right represents MLG, the adjacent arrows represent the spin orientations of the nearest layers of $\rm CrI_3$, and the leftmost arrow represents the $\rm CrI_3$ layer furthest from MLG. When the magnetic field is less than 0.8 T, $\rm CrI_3$ exhibits AFM coupling. However, around 0.8 T, the spins of the top two layers of $\rm CrI_3$ transition to FM with a spin-up orientation, and around 1.8 T, all layers of $\rm CrI_3$ transition to a spin-up orientation. However, the shift at 0.8 T is much weaker than at 1.8 T, indicating that graphene is more sensitive to the coupling between the nearest two layers of $\rm CrI_3$.

\begin{figure*}[t]

\includegraphics[scale=0.5]{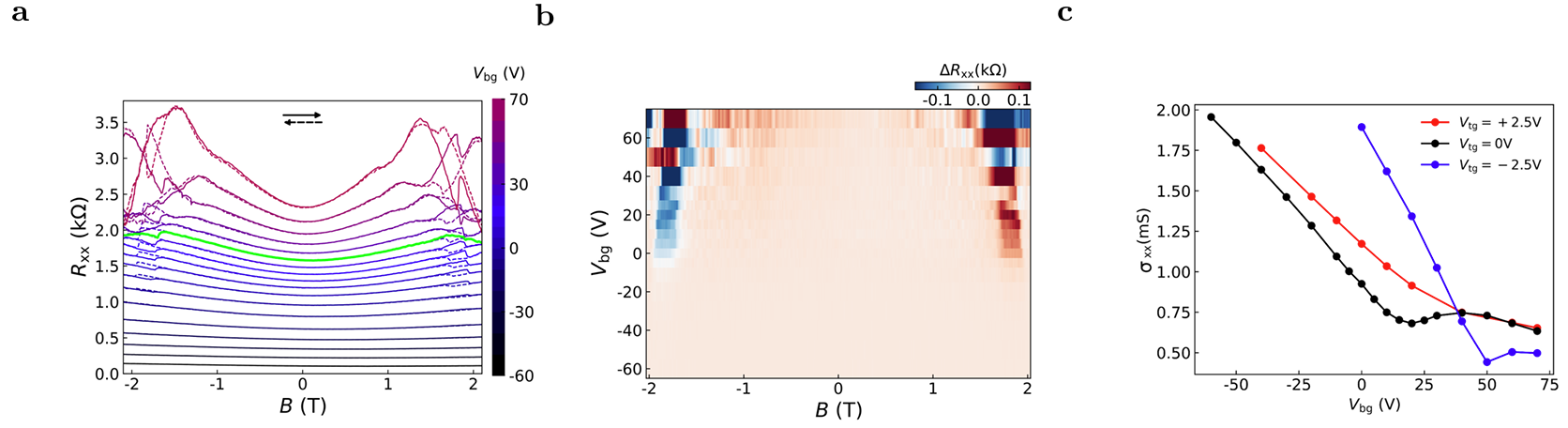}
\caption{\textbf{Resistance loops of $\rm CrI_3$/MLG.} \textbf{a.} $R_{\rm xx}$ acquired by sweeping $B$ forth and back at different values of $V_{\rm bg}$ with $V_{\rm tg}=$0 V. All the curves have been offset for clarity. \textbf{b.} Color map of $\Delta R$ in $B$ and $V_{\rm bg}$ parameter space, $\Delta R$  extracted by the difference in resistance during the back-and-forth sweeping of $B$. \textbf{c.} The conductance $\sigma_{\rm xx}$  varies with $V_{\rm bg}$ for $V_{\rm tg}$ fixed at three different values and $B=0$ T.}
\end{figure*}

\label{Fig 3 loop B}
Sweeping the magnetic field forth and back, helps to reveal the shifts in resistance at $B_{\rm c}$ more clearly. Fig 3a shows the magnetic field sweeping with $V_{\rm tg}$ fixed at 0 V and $V_{\rm bg}$ ranging from -60 V to 70 V. All the curves have been offset for clarity. The horizontal black arrows represent the sweeping direction. When $V_{\rm bg}$ is larger than 10 V, there is a distinct loop in $R_{\rm xx}$ at $|B|\approx 1.8$ T. $V_{\rm bg}$ = 10 V coincides with the $R_{\rm AP}$ corresponding gate value in Fig 1d. The experimental results show that the loops around $B_{\rm c}$ can be clearly observed in region II. This can be explained as follows: in region I, the conduction band minimum of $\rm CrI_3$ is higher than the Fermi level, which inhibits charge transfer. Therefore, MLG cannot perceive the difference between AFM and FM states.

Fig 3a shows that the amplitude of the loops decreases at $V_{\rm bg}=$30 V (highlighted in bright green). Additionally, during the switching of the magnetic configuration, the resistance jumps down and up. This is because the energy band of MLG is composed of split Landau levels, as shown in Fig 4d. $\varepsilon_{F}$ represents the position of the Fermi energy before the shift, and $\varepsilon_{F}^{'}$ represents the Fermi energy after the shift. If $\varepsilon_{F}^{'}$  is between $\rm B_0$ and $\rm B_1$, the carrier density increases after the shift, causing the resistance to jump down. If it is precisely at $\rm B_1$, the carrier density remains almost unchanged, resulting in the near disappearance of the loop. If it is between $\rm B_1$ and $\rm B_2$, the resistance jumps due to the decrease in carrier density.

Fig 3b shows the variation of $\Delta R_{\rm xx}$ with $B$ and $V_{\rm bg}$, where $\Delta R_{\rm xx}=R_{\rm xx-forth}-R_{\rm xx-back}$. $\Delta R_{\rm xx}$ is antisymmetric with respect to the magnetic field. This indicates that the bottom of the conduction bands for the FM states  ($\uparrow\uparrow$ and $\downarrow\downarrow$) and the AFM states ($\uparrow\downarrow$ and $\downarrow\uparrow$) are energetically degenerate, respectively. However, as $V_{\rm bg}$ increases, $B_{\rm c}$ gradually decreases. This is because the critical magnetic field of $\rm CrI_3$ decreases with increasing doping, which is consistent with recent studies on doping-controlled interlayer coupling in $\rm CrI_3$\cite{Huang2018NatureNanotechElectricalControl2Da, Jiang2018NatureNanotechControllingMagnetism2D}. The critical temperature of $\rm CrI_3$ is  also reduced due to charge doping (Fig S6).

\label{Fig 4band}
\begin{figure*}[t]
	\includegraphics[scale=0.5]{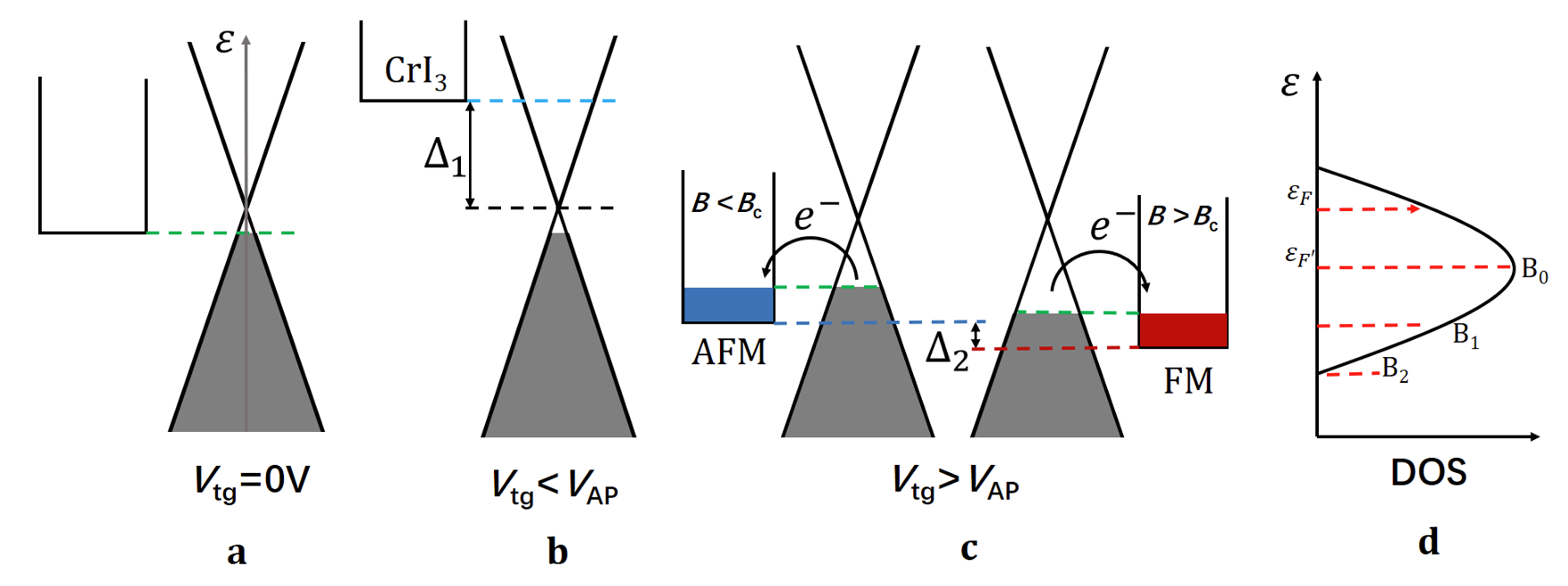}
	\caption{\textbf{a.} The schematic energy band diagram of $\rm CrI_3$ (rectangular) and MLG (conical) without gate control. \textbf{b.} Under gate control, the conduction band of $\rm CrI_3$ moves upwards, crossing over the Dirac point. \textbf{c.} Under gate control, the conduction band of $\rm CrI_3$ moves downwards, below the Fermi energy of MLG, and gets filled with electrons. \textbf{d.} A schematic diagram depicts the shift in resistance near $B_{\rm c}$.}
\end{figure*}
Fig 3c shows the relationship between conductivity $\sigma$ and $V_{\rm bg}$ at zero magnetic field, where the conductance is calculated by the formula $\sigma_{\rm xx}=\rho_{\rm xx}/(\rho_{\rm xx}^2+\rho_{\rm xy}^2)$, where $\rho_{\rm xx}=R_{\rm xx}\times W/L$ and $\rho_{\rm xy}=R_{\rm xy}$, $W,L$ are the width and the length of the channel, respectively. In region I, the conductivity varies linearly with $V_{\rm bg}$. However, in region II, the conductivity no longer exhibits a linear relationship with $V_{\rm bg}$. Even within region I, as $V_{\rm tg}$ increases, the slope decreases continuously. Combining $\sigma_{\rm xx}\propto |e|n_{\rm g}\mu$ and $n_{\rm g}\propto C_{\rm bg}(V_{\rm bg}-V_{\rm bg0})/e$, where $e$ is the elementary charge, $n_{\rm g}$ is the concentration, $\mu$ is the mobility, and $V_{\rm bg0}$ is the back gate voltage corresponding to the Dirac point, we can calculate $|\partial \sigma_{\rm xx}$/$\partial V_{\rm bg}|\sim\mu $. The reduced slope (i.e. the slope of $\sigma_{\rm xx}$ against $V_{\rm bg}$) indicates that the mobility is decreasing which is relevant to the scattering on remote magnetic impurities in $\rm CrI_3$. The temperature-dependent resistance in $\rm CrI_3$/graphene is also anomalous, as shown Fig S7. In ordinary graphene, the resistance decreases with decreasing temperature, but in the $\rm CrI_3$/graphene region, the resistance increases with decreasing temperature. This is also due to the impurities introduced by $\rm CrI_3$, which result in additional scattering effects in graphene apart from phonon scattering.

A proximity exchange field has been observed unambiguously in transport measurements of $\rm WT_2/CrI_3$ \cite{Zhao2020NatMaterMagneticProximityNonreciprocal}and optical spectroscopy measurements of $\rm WSe_2/CrI_3$\cite{Mukherjee2020NatCommunObservationSitecontrolledLocalized}. In $\rm CrBr_3$/graphene, a magnetoresistance exhibits hysteresis in the in-plane magnetic field around zero, indicating the presence of an exchange field\cite{Ghazaryan2018NatElectronMagnonassistedTunnellingVan}. $\rm CrBr_3$ is a ferromagnetic material, making it more susceptible to achieving a magnetic proximity effect. Currently, there is greater experimental focus on the vertical magnetic field, with relatively less attention given to the parallel magnetic field. It is not clear for the unexpectedly small exchange coupling for graphene on $\rm CrI_3$. We speculate on the possible two reasons for the small exchange coupling. One is the band structures of $\rm {CrI_3}$ and graphene, which leads to a significant charge transfer effect at the interface, and the transferred electrons mainly occupy the non-spin-polarized iodine (I) atoms. These electrons will screen chromium (Cr) atoms and graphene coupling\cite{Ren2023AdvancedPhysicsResearchTunableQuantumAnomalous}. The other one is the  impurities in $\rm {CrI_3}$, as indicated by the anomalous temperature dependence,  and significantly reduces the mobility of graphene, making the observation of possible anomalous Hall effect more difficult.

In summary, the band alignment in graphene/$\rm CrI_3$ heterostructures is gate-tunable. Charge carriers can be transferred between graphene and $\rm CrI_3$ under the control of gate voltages and even be suppressed entirely. This results in significantly different field-effect behavior in graphene, as the gate voltage no longer monotonically injects electrons into graphene. The electrons will either accumulate in graphene or transfer to $\rm CrI_3$ based on the band alignment and the density of states in graphene. The electrons accumulated in $\rm CrI_3$ will screen the gate voltage on that side. Additionally, due to the varying band structure of $\rm CrI_3$ under different magnetic couplings, the bottom of the conduction band in FM-coupling states has a smaller value. Therefore, by switching the interlayer coupling of $\rm CrI_3$ through a magnetic field, the resistance of graphene exhibits loops at the critical magnetic field. Through this phenomenon, we discovered that the critical magnetic field and critical temperature of $\rm CrI_3$ decrease with increasing doping. However, the specific mechanism of the magnetic proximity effect between $\rm CrI_3$ and graphene for anomalous Hall effect still requires further research.

\section*{Supplementary Material}
Atomic force microscopy images; details of the device fabrication; summary of devices; quantitative estimation of the density of charge transfer and the graphene mobility;  transport in a device with a thin hBN spacer; temperature-dependent transport.

\section*{Acknowledgements}
We thank S. Jiang, G. Ma, D. Zhang and J. Huang for their fruitful discussions.
Financial support from the National Key R\&D  Program of China (2018YFA0306800, Nos. 2018YFA0305804), the National Natural Science Foundation of China (Nos. 12004173, 11974169), the Natural Science Foundation of Jiangsu Province (Nos. BK20220066), and the Fundamental Research Funds for the Central Universities (Nos. 020414380087, 020414913201) are gratefully acknowledged.

\subsection*{Conflict of Interest}
The authors have no conflicts to disclose.

\section*{Data Availability}
The data supporting this study's findings are available from the corresponding author upon reasonable request.

\bibliographystyle{iopart-num}
\bibliography{Refs}

\end{document}